\documentclass[preprint,aps,prl,showpacs,twocolumn,footinbib,10pt]{revtex4}
\usepackage{epsfig,epsf,graphics}
\usepackage{amsfonts,amssymb,amsmath}
\usepackage{color}

\newcommand{\be}{\begin{equation}}
\newcommand{\ee}{\end{equation}}
\newcommand{\ba}{\begin{eqnarray}}
\newcommand{\ea}{\end{eqnarray}}
\newcommand{\non}{\nonumber}

\newcommand{\al}{&\!\!\!}

\begin{document}

\preprint{\small FZJ-IKP-TH-2011-2}

\title{More kaonic bound states and a comprehensive interpretation of the $D_{sJ}$ states}

\author{Feng-Kun Guo$^1$}
      \email{fkguo@hiskp.uni-bonn.de}
\author{Ulf-G. Mei{\ss}ner$^{1,2}$}
      \email{meissner@hiskp.uni-bonn.de}
\affiliation{$\rm ^1$Helmholtz-Institut f\"ur Strahlen- und
             Kernphysik and Bethe Center for Theoretical Physics,\\ Universit\"at
             Bonn,  D--53115 Bonn, Germany}
\affiliation{$\rm ^2$Institut f\"{u}r Kernphysik, J\"ulich Center
          for Hadron Physics and Institute for Advanced Simulation, Forschungszentrum J\"{u}lich,
             D--52425 J\"{u}lich, Germany}%


\begin{abstract}
The leading order interaction between a Goldstone boson and a matter field is
universally dominated by the Weinberg-Tomozawa term. Based on this observation,
we predict a rich spectrum of bound states of a kaon and a heavy meson. We argue
that if the life time of an excited heavy meson is significantly longer than the
range of forces, then the finite width of that state can be neglected in a first
approximation. Then, the $D_{s0}^*(2317)$, $D_{s1}(2460)$, $D_{sJ}(2860)$ and
$D_{sJ}(3040)$ are generated as $DK$, $D^*K$, $D_1(2420)K$ and $D^*(2600)K$
bound states, respectively. In addition to the remarkable agreement with the
measured masses, the decay patterns of the $D_{sJ}(2860)$ and $D_{sJ}(3040)$ can
also be understood. Two more $D_{sJ}$ states, and kaonic bound states with the
bottom mesons as well as the doubly charmed baryon are also predicted.
\end{abstract}






\maketitle



The leading order (LO) interaction between a Goldstone boson and a matter field
is universally dominated by the Weinberg-Tomozawa (WT)
term~\cite{Weinberg:1966kf,Tomozawa:1966jm}. With the kaon being a Goldstone
boson of the spontaneous chiral symmetry breaking, the $D_{s0}^*(2317)$ and
$D_{s1}(2460)$ can be generated as bound states of the $DK$ and $D^*K$ systems,
respectively~\cite{Kolomeitsev:2003ac,Guo:2006fu,Guo:2006rp,Gamermann:2006nm,Guo:2008gp,Guo:2009ct,Cleven:2010aw}.
Considering a heavy matter field $H$ with a mass $M_H$ much larger than the kaon
mass $M_K$, the heavy field can be treated nonrelativistically. Then the LO
$S$-wave elastic scattering amplitude from the WT term can be given in a simple
form,
\be%
V(s) = C\frac{M_H E_l(s)}{F_\pi^2}, \label{eq:vlodk}
\ee%
where $s$ is the squared energy in the center-of-mass frame, $E_l$ is the energy
of the light meson in the center-of-mass frame, and $F_\pi=92.4$~MeV is the pion
decay constant. The coefficient $C=-2$ for the $I=0$ $HK\to HK$ channel. The
heavy matter fields for which Eq.~(\ref{eq:vlodk}) can be applied include the
excited heavy mesons with flavor contents $c\bar q,b\bar q$, and the doubly
heavy antibaryons $\bar c\bar c \bar q$ and $\bar b\bar b \bar q$. (A subtlety
related to the finite width effects will be discussed later.) Therefore, if the
$D_{s0}^*(2317)$ is a $DK$ bound state as suggested in
Refs.~\cite{Barnes:2003dj,Kolomeitsev:2003ac,Guo:2006fu,Gamermann:2006nm}, one
can easily expect that there must be more bound states. In fact, the $\bar BK$
and $\bar B^*K$  bound states have been
predicted~\cite{Kolomeitsev:2003ac,Guo:2006fu,Guo:2006rp,Cleven:2010aw}. Here a
bound state does not strictly mean a state with a zero width. It means if all
the other particles are neglected, then it has a vanishing width. If some other
particles with lower masses are switched on as in the real world, these ``bound
states'' will have a finite width.

On the experimental side, besides the $D_{s0}^*(2317)$ and $D_{s1}(2460)$, more
charmed strange mesons have been discovered in recent years. They include the
$D_{s1}^*(2700)$~\cite{Aubert:2006mh,Brodzicka:2007aa},
$D_{sJ}(2860)$~\cite{Aubert:2006mh} and $D_{sJ}(3040)$~\cite{Aubert:2009di}.
Both the $D_{s1}^*(2700)$ and the $D_{sJ}(2860)$ decay into $DK$ and $D^*K$,
while the $D_{sJ}(3040)$ was only observed in the $D^*K$ final
state~\cite{Aubert:2009di}. The ratios between different decay modes of the
$D_{sJ}(2860)$ were measured to be~\cite{PDG2010}
\ba%
R_{D_{sJ}(2860)} = \frac{\Gamma(D_{sJ}(2860)\to D^*K)}{\Gamma(D_{sJ}(2860)\to
DK)} = 1.10\pm0.24,
\ea%
and a similar ratio for the $D_{s1}^*(2700)$ is
$R_{D_{s1}^*(2700)}=0.91\pm0.18$. Both states have a natural parity, i.e. a
positive (negative) parity for an even (odd) spin, because they decay into the
$DK$ channel. Using the LO heavy hadron chiral perturbation theory (HHChPT),
which is model-independent, the decay pattern of the $D_{s1}^*(2700)$ for a
$(2S, J^P=1^-)$ assignment
was calculated to be $R_{D_{s1}^*(2700)}=0.91\pm0.04$~\cite{Colangelo:2007ds},
fully consistent with the data. Furthermore, the mass of the $D_{s1}^*(2700)$
agrees well with the prediction in the Godfrey-Isgur quark
model~\cite{Godfrey:1985xj}. So there is little doubt that the $D_{s1}^*(2700)$
is the $(2S, 1^-)$ $c\bar s$ meson. The situation for the $D_{sJ}(2860)$ and
$D_{sJ}(3040)$ is not clear yet. Their quantum numbers are not known. The LO
results from the HHChPT for $R_{D_{sJ}(2860)}$ were given
in~\cite{Colangelo:2006rq} for different assignments. They are 1.23, 0, 0.63,
0.06, and 0.39 for $(2S, 1^-)$, $(2P, 0^+)$, $(2P, 2^+)$, $(1D, 1^-)$, and $(1D,
3^-)$, respectively. Comparing with the measured value, one can conclude that
the only possibilities are $(2S, 1^-)$ and $(2P, 2^+)$. However, the $(2S, 1^-)$
$c\bar s$ meson has already been identified as the $D_{s1}^*(2700)$, and the
$(2P, 2^+)$ one would have a mass as large as 3.1~GeV~\cite{Di Pierro:2001uu}.
Various explanations of the $D_{sJ}(2860)$ and $D_{sJ}(3040)$ have already been
discussed in the
literature~\cite{vanBeveren:2006st,Close:2006gr,Zhang:2006yj,Vijande:2008zn,Zhong:2009sk,Ebert:2009ua,
Colangelo:2010te,Chen:2009zt,Sun:2009tg}.

In this paper, we will show that all of the $D_{s0}^*(2317)$, $D_{s1}(2460)$,
$D_{sJ}(2860)$ and $D_{sJ}(3040)$ mesons can be interpreted as charmed-meson
kaon bound states. The $D_{sJ}(2860)$ and $D_{sJ}(3040)$ are interpreted as
$D_1(2420)K$ and $D^*(2600)K$ bound states, respectively. In addition, more
kaonic bound states will be predicted.

One important issue requires discussion. Both the $D$ and $D^*$ have a
negligible width, however, the excited heavy mesons normally have a width of
tens of MeV or even larger than 100~MeV. The finite width presents another
energy scale which might invalidate the use of the WT term. A natural question
is: how large a width can be neglected for a given case? Let us consider a
hadron with a finite life time $\tau$. If $\tau$ is long enough
so that the interaction responsible for the binding takes place, then one can
neglect the width in calculating the bound state masses. Mathematically, this
means $\tau\gg r$, with $r$ the range of forces, i.e. $\Gamma\ll1/r$. For the
interaction between a kaon and a heavy meson, since a pion cannot be exchanged,
the range of forces is set by $1/r\sim M_\rho$ or more safely by $2M_\pi$.
Therefore, if the width of the other hadron is much smaller than $M_\rho$, then
in a first approximation, the width can be neglected. These heavy mesons,
written in terms of $\{H,H_s\}$ in the same SU(3) triplet, include $\{D,D_s\}$,
$\{D^*,D_s^*\}$, $\{D_1(2420)$, $D_{s1}(2536)\}$, $\{D_2^*(2460),
D_{s2}^*(2573)\}$, $\{D(2550), ?\}$, $\{D^*(2600)$, $D_{s1}^*(2700)\}$, and
$\{\bar B, \bar B_s\}$, $\{\bar B^*, \bar B_s^*\}$, $\{B_1(5720)$,
$B_{s1}(5830)\}$, $\{B_2^*(5747), B_{s2}(5840)\}$~\cite{PDG2010}, and their
strange partners. Among them, the $D(2550)$ and $D^*(2600)$ were discovered very
recently by the BABAR Collaboration~\cite{delAmoSanchez:2010vq}. The properties
of the $D^*(2600)$ are consistent with the radially excited $(2S,1^-)$ charmed
meson. The $D(2550)$ might be the $(2S,0^-)$ state since it has the correct
angular distribution~\cite{delAmoSanchez:2010vq}. However, its width is much
larger than the prediction using the LO HHChPT,
$\Gamma_{D(2550)}=0.55\Gamma_{D^*(2600)}\sim50\,$~MeV~\cite{inpre}.
Nevertheless, we will tentatively assume the $D(2550)$ as the $2S$ pseudoscalar
charmed meson.
On the contrary, however, this is not the case for the $\bar D_1(2430)D^*$
system which was considered in Ref.~\cite{Close:2009ag}. The width of the
$D_1(2430)$ is $\sim400$~MeV $\gg M_\pi$, whose inverse sets the range of forces
for this system, and hence the finite width effect cannot be neglected. Indeed,
it was shown that, after considering the finite width or the three-body cut, the
bound state disappears~\cite{Filin:2010se}. The impact of the finite width
effect on the line shape of a composite particle was considered in
Ref.~\cite{Hanhart:2010wh} (see also \cite{Artoisenet:2010va}).

\begin{table*}[t]
\caption{\label{tab:charm}Predicted masses of charmed-meson kaon bound states.
The experimental values are listed in the fourth row for a comparison. The
expected dominant decay modes are also given.}
\begin{ruledtabular}
\begin{tabular}{ l | c  c c c c c }
 Main constituents & $DK$  &  $D^*K$ & $D_1(2420)K$ & $D_2(2460)K$ & $D(2550)K$ & $D^*(2600)K$ \\\hline
 $J^P$ & $0^+$ & $1^+$ & $1^-$ & $2^-$ & $0^+$ & $1^+$\\
 Predicted masses & 2317.8(input) & $2458\pm3$ & $2870\pm9$ & $2910\pm9$ & $2984\pm10$ & $3052\pm11$ \\
 Experimental data & $2317.8\pm0.6$ & $2459.5\pm0.6$ & $2862\pm2^{+5}_{-2}$ & & &
 $3044\pm8^{+30}_{-5}$\\
 Dominant decays & $D_s\pi$ & $D_s^*\pi$ & $D^{(*)}K,D_s^{(*)}\eta$ & $D^*K,D_s^*\eta$ & $DK,D_s\eta$ & $D^*K,D_s^*\eta,D_s\omega,DK^*,D\phi$ \\
\end{tabular}
\end{ruledtabular}
\end{table*}
The masses of the bound states can be calculated by searching for poles in the
resummed $S$-wave $I=0$ scattering amplitudes $T(s) =
V(s)[1-G(s)V(s)]^{-1}$~\cite{Oller:1997ti,Oller:2000fj}. We will consider two
coupled channels for each case, which means all of $T(s),V(s)$ and $G(s)$ are
$2\times2$ matrices. Denoting the nonstrange (strange) heavy meson by $H_{(s)}$,
the matrix elements of the symmetric matrix $V(s)$ is given by
Eq.~(\ref{eq:vlodk}). For $I=0$, the coefficient $C=-2,0$ and $-\sqrt{3}$ for
$V_{HK\to HK}(s),V_{H_s\eta\to H_s\eta}(s)$ and $V_{HK\to H_s\eta}(s)$,
respectively. $G(s)$ is a diagonal matrix with the nonvanishing elements given
by the loop function for a nonrelativistic heavy meson and a relativistic light
meson~\cite{Cleven:2010aw}
\begin{eqnarray}\nonumber
G_{Hl}(s)\al=\al
\frac{1}{16\pi^2
  M_H}\bigg\{E_l\left[a(\mu){+}\log\left(\frac{M_l^2}{\mu^2}\right)\right] \non\\
\al\al + 2|\vec p_l|\cosh^{-1}\left(\frac{E_l}{M_l}\right) - 2\pi
i|\vec p_l| \bigg\},
\label{eq:NRloop}
\end{eqnarray}
with $a(\mu)$ a subtraction constant, $\mu$ the scale of dimensional
regularization, and $\vec p_l (M_l)$ the three-momentum (mass) of the light
meson. Equation~(\ref{eq:vlodk}) seems to imply that the heavier the matter
field is, the stronger the interaction will be. However, the factor $M_H$ will
be canceled by a factor of $1/M_H$ in the loop function in the resummed
amplitudes, as required by the heavy quark symmetry. It is natural to choose the
same value of the subtraction constant $a(\mu)$ for all the heavy hadrons. A
change in the scale $\mu$ can be balanced by a corresponding change in $a(\mu)$.
However, for checking the stability of the results, we will also fix the
subtraction constant while allowing $\mu$ varying from $1$~GeV to $M_H$. For a
given $\mu$, $a(\mu)$ is fixed by reproducing the mass of the $D_{s0}^*(2317)$,
and we get $a(1~{\rm GeV})=-3.84$. The results for the charmed mesons are
summarized in Table~\ref{tab:charm} with the expected dominant decay modes. We
use the central values for the masses of the constituents. The error bars
reflect the variation of $\mu$. It has been assumed that the unknown strange
partner of the $D(2550)$ has a mass 100~MeV heavier than the $D(2550)$. In fact,
neglecting the $H_s\eta$ channel and keeping only the $HK$ channel, the results
are almost the same, with a difference within 4~MeV. This means one can
interpret the generated states as heavy-meson kaon bound states.
We can see that with input only  from the $D_{s0}^*(2317)$, the masses of the
$D_{s1}(2460)$, $D_{sJ}(2860)$ and $D_{sJ}(3040)$ can be well reproduced. In the
following, we will focus on the latter two states, which are interpreted as
$S$-wave $D_1(2420)K$ and $D^*(2600)K$ bound states, respectively. In this
scenario, the $J^P$ are $1^-$ for the $D_{sJ}(2860)$, and $1^+$ for the
$D_{sJ}(3040)$.
We also checked that, using $F_K$ instead of $F_\pi$ in Eq.~(\ref{eq:vlodk}),
which represents some of the higher order corrections in the chiral expansion,
the results are very stable. The predicted masses change for no more than 3~MeV
with the subtraction constant refitted to the mass of the $D_{s0}^*(2317)$.
Inclusion of other coupled channels such as those listed as decay modes in
Table~\ref{tab:charm} involves unknown coupling constants. As long as the
threshold of the coupled channel $T_{\rm CC}$ is far from the bound state mass,
which can be characterized as $|T_{\rm CC}-M_{\rm BS}|\gg \epsilon$, with
$\epsilon$ being the binding energy, they will modify the mass only marginally.
However, their presence will give a finite width to the bound state. Now let us
discuss their decay modes.

\begin{figure}[tb]
\begin{center}
\vglue-0mm
\includegraphics[width=0.49\textwidth]{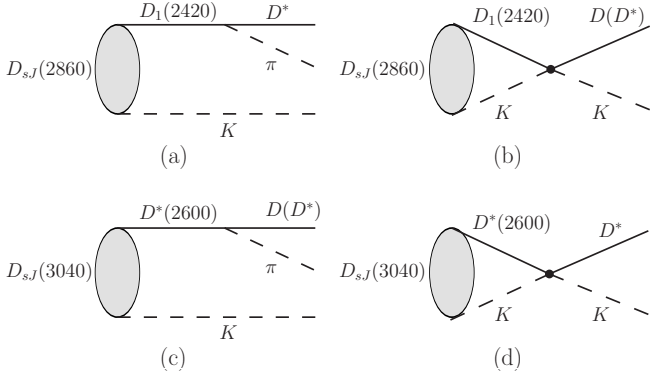}
\vglue-0mm \caption{Schematic diagrams of some of the decay mechanisms (not a
complete list) of the $D_{sJ}(2860)$ and $D_{sJ}(3040)$. \label{fig:decays}}
\end{center}
\vspace{-0.45cm}
\end{figure}
Some of the decay mechanisms of the  $D_{sJ}(2860)$ and $D_{sJ}(3040)$ are shown
in Fig.~\ref{fig:decays}. Note that this is not a complete list, for instance,
the $D_s\omega$ is another important decay channel of the $D_{sJ}(3040)$. In
fact, this channel can be used in distinguishing the $D^*(2600)K$ bound state
picture from a $c\bar s$ picture, for which the decay into $D_s\omega$ would be
very small due to the OZI suppression. Being a $1^+$ state, the $D_{sJ}(3040)$
cannot decay into the $DK$. This is consistent with the experimental facts. The
two-body decays of the $D_{sJ}(2860)$ and $D_{sJ}(3040)$ occur through the
scattering $D_1(2420)K\to D^{(*)}K$ and $D^*(2600)K\to D^*K$, respectively, as
shown in Fig.~\ref{fig:decays}. The total angular momentum of the light degrees
of freedom in a heavy hadron $s_\ell$ is conserved in the heavy quark limit. The
axial-vector meson $D_1(2420)$ has $s_\ell^P=\frac{3}{2}^+$. So without breaking
the heavy quark spin symmetry, the $D_1(2420)K$ can couple to the $DK$ and
$D^*K$ channels in a $P$ wave because $D^{(*)}$ has $s_\ell^P=\frac{1}{2}^-$.
Therefore, the two-body decay modes $D^{*}K$ and $DK$ can be related through
heavy quark spin symmetry. Nonrelativistically, the decay amplitudes can be
written as
\ba%
\label{eq:A2860} {\cal M}(D_{sJ}(2860)\to DK) \al=\al g_{D} G_{D_1K}
\vec{\varepsilon}_{D_{sJ}}\cdot\vec{k}_D, \non\\
{\cal M}(D_{sJ}(2860)\to D^*K) \al=\al g_{D^*} G_{D_1K}\epsilon^{ijk} k^i_{D^*}
\varepsilon_{D_{sJ}}^j \varepsilon^{*k}_{D^*},
\ea%
where $G_{D_1K}$ is the $D_1K$ loop function, $\vec{k}_{D^{(*)}}$ is the
momentum of the $D^{(*)}$ in the center-of-mass frame, $\epsilon^{ijk}$ is the
antisymmetric Levi-Civita tensor, and $\vec{\varepsilon}_{D_{sJ}}$ and
$\vec{\varepsilon}_{D^*}$ are the polarization vectors of the $D_{sJ}(2860)$ and
$D^*$, respectively.
The heavy quark spin symmetry requires $g_D/g_{D^*}=\sqrt{M_D/M_{D^*}}$
taking into account the nonrelativistic normalization factor. The result
\be%
\label{eq:2860ratio} R_{D_{sJ}(2860)} = 2 \frac{M_{D^*}}{M_D}
\left|\frac{\vec{k}_{D^*}}{\vec{k}_D}\right|^3 = 1.23
\ee%
is in  nice agreement with the data $1.10\pm0.24$. Furthermore, the helicity
angular distribution for the sequential process $D_{sJ}(2860)\to D^*K\to D\pi K$
derived from Eq.~(\ref{eq:A2860}) is $\sin^2\theta$, which again agrees with the
observation~\cite{Aubert:2009di}.

One expects that the $D_{sJ}(2860)$ and $D_{sJ}(3040)$ can also decay through
the decays of the $D_1(2420)$ and $D^*(2600)$, respectively. However, the
partial widths of these sequential decays are not large. This is because they
are suppressed by a $D$-wave and $P$-wave factor for the $D_{sJ}(2860)$ and
$D_{sJ}(3040)$, respectively, as well as by the three-body phase space. In fact,
this expectation can be confirmed by a rough estimate. For an $S$-wave loosely
bound state, the binding energy $\epsilon$ and effective coupling $g$ for the
bound state to its constituents are related by
$g^2=16\pi(m_1+m_2)^2\sqrt{2\epsilon/\mu}[1+{\cal
O}(r\sqrt{2\mu\epsilon})]$~\cite{molecule1,molecule2}, where $m_{1,2}$ are
the masses of the constituents, and $\mu$ 
is their reduced mass. We get 1 and 26~MeV for such sequential decays of the
$D_{sJ}(2860)$ and $D_{sJ}(3040)$, respectively, which means they only
contribute a branching fraction of about $2\%$ and $10\%$, respectively.
However, these results can only be regarded as an order-of-magnitude estimate,
since for both cases, $\sqrt{2\mu\epsilon}\simeq220$~MeV, and so the uncertainty
in the so-estimated effective coupling is very large. Nevertheless, the two-body
decays would be the dominant modes for these two states. Therefore, one expects
$\Gamma_{D_{sJ}(3040)}\gg\Gamma_{D_{sJ}(2860)}$, consistent with the data,
because the two-body decays of the $D_{sJ}(3040)$ and $D_{sJ}(2860)$ are in an
$S$- and $P$-wave, respectively.

In addition, heavy quark spin symmetry implies that each of the predicted bound
states has its spin multiplet partner~\cite{Guo:2009id}. For instance, the
$D_{s1}(2460)$ is the spin partner of the $D_{s0}^*(2317)$. Because the kaon,
which has a negative parity, interacts with the heavy meson in an $S$-wave, so
for the $D^{(*)}K$ systems, $s_\ell^P=\frac12^+$. Note that they should not be
confused with the $s_\ell^P=\frac12^+$ $c\bar s$ mesons because they have a
different dynamical origin. Similarly, the $D_2(2460)K$ bound state, to be
called $D_{s2}^*(2910)$, is the spin partner of the $D_{sJ}(2860)$, and they
have $s_\ell^P=\frac32^-$. The $D_{sJ}(3040)$ and the $D(2550)K$ bound state, to
be called $D_{s0}^*(2985)$, form another $s_\ell^P=\frac12^+$ doublet. They can
be regarded as the excited states of the $D_{s0}^*(2317)$ and $D_{s1}(2460)$.

Assuming the decay modes given in Table~\ref{tab:charm} exhaust the decay
widths, ratios of the total widths can be predicted based on the HHChPT. At LO,
we obtain
\be%
\frac{\Gamma_{D_{s2}^*(2910)}}{\Gamma_{D_{sJ}(2860)}} \simeq 0.2, \qquad
\frac{\Gamma_{D_{s0}^*(2985)}}{\Gamma_{D_{sJ}(3040)}} < 1,
\ee%
where the less-than sign is because the $DK^*,D_s\omega,D_s\phi$ channels for
the $D_{sJ}(3040)$ were not taken into account. Thus we have
$\Gamma_{D_{s2}^*(2910)}\sim10\,$~MeV. Such a small width suggests that more
experimental efforts should be devoted to the $D^*K$ data.

We can make further predictions for the bottom sector. The results are listed in
Table~\ref{tab:bottom} together with the expected dominant decay channels.
\begingroup
\squeezetable
\begin{table}[th]
\caption{\label{tab:bottom}Predicted masses of bottom-meson kaon bound states.
The expected dominant decay modes are also given.}
\begin{ruledtabular}
\begin{tabular}{ l | c  c c c }
 Constituents & $\bar BK$ & $\bar B^*K$ & $\bar B_1(5720)K$ & $\bar B_2(5747)K$
 \\\hline
 $J^P$ & $0^+$ & $1^+$ & $1^-$ & $2^-$\\
 Predicted masses & $5705\pm31$ & $5751\pm32$ & $6151\pm33$ & $6169\pm33$  \\
 Dominant decays & $\bar B_s\pi$ & $\bar B_s^*\pi$ & $\bar B^{(*)}K,\bar B_s^{(*)}\eta$ & $\bar B^*K,\bar B_s^*\eta$\\
\end{tabular}
\end{ruledtabular}
\end{table}
\endgroup
In addition, there should also be kaonic bound states with the doubly heavy
baryons. For instance, the $\bar\Xi_{cc}(3520)K$ bound state is predicted to
have a mass of $3956\pm20$~MeV.

In summary, using  chiral and heavy quark symmetry, we predicted a rich spectrum
of kaonic bound states. We argue that, if the lifetime of the constituents of a
bound state is significantly longer than the range of forces, then the finite
width effect can be neglected in a first approximation. The $D_1(2420)K$ and
$D^*(2600)K$ bound states are found to fit very well to both the measured masses
and decay patterns for the $D_{sJ}(2860)$ and $D_{sJ}(3040)$, respectively.
Hence, all known properties of the $D_{s0}^*(2317)$, $D_{s1}(2460)$,
$D_{sJ}(2860)$ and $D_{sJ}(3040)$ can be systematically explained as kaonic
bound states. Kaonic bound states for the bottom mesons and doubly charmed
baryon are also predicted. The decay mode $D_{sJ}(3040)\to D_s\omega$ can serve
as a criterion in distinguishing the present interpretation from a $c\bar s$
picture. The narrow $D_{s2}^*(2910)$ and broader $D_{s0}^*(2985)$  should be
searched for in the $D^*K$ and $DK$ channels, respectively. In order to
understand the low-energy strong interaction better, searching for these states
should be an important experimental issue.

\medskip

\begin{acknowledgments}
We are grateful to C. Hanhart for useful comments. We thank the HGF for funds
provided to the virtual institute ``Spin and strong QCD'' (VH-VI-231), the DFG
(SFB/TR 16) and the EU I3HP ``Study of Strongly Interacting Matter'' under the
Seventh Framework Program of the EU. U.-G. M. also thanks the BMBF for support
(Grant No. 06BN9006).
\end{acknowledgments}


\end{document}